# Superconductivity and Phase Diagram in Iron-based Arsenic-oxides ReFeAsO$_{1-\delta}$ (Re = rare earth metal) without Fluorine Doping

Zhi-An Ren*, Guang-Can Che, Xiao-Li Dong, Jie Yang, Wei Lu, Wei Yi, Xiao-Li Shen, Zheng-Cai Li, Li-Ling Sun, Fang Zhou, Zhong-Xian Zhao*

National Laboratory for Superconductivity, Institute of Physics and Beijing National Laboratory for Condensed Matter Physics, Chinese Academy of Sciences, P. O. Box 603, Beijing 100190, P. R. China

Here we report a new class of superconductors prepared by high pressure synthesis in the quaternary family ReFeAsO$_{1-\delta}$ (Re = Sm, Nd, Pr, Ce, La) without fluorine doping. The onset superconducting critical temperature ($T_c$) in these compounds increases with the reduction of Re atom size, and the highest $T_c$ obtained so far is 55 K in SmFeAsO$_{1-\delta}$. For the NdFeAsO$_{1-\delta}$ compound with different oxygen concentration a dome-shaped phase diagram was found.

While the underlying physical mechanism for high-$T_c$ superconductivity (SC) in copper oxides discovered 22 years ago [1] is still under debate, some iron- (nickel-) based oxypnictide superconductors were discovered [2, 3]. Following the recent report of $T_c \sim 26$ K in fluorine (F)-doped arsenic-oxide LaFeAs[$O_{1-x}F_x$] [4], extensive research efforts have been devoted to this system due to the relative high $T_c$, the two-dimensional layered structure like cuprates and the presence of iron, usually correlated with ferromagnetism. The $T_c$ has been very rapidly enhanced to above 50 K by substituting La with smaller Re elements, putting these materials to another high-$T_c$ family [5-9].

These quaternary equiatomic ReFeAsO compounds have a rather simple structure of alternating Fe-As and Re-O layers with eight atoms in a tetragonal unit cell (space group: P4/nmm, similar to that of ReCuSeO in Ref. [10]), where the Fe-As layer is thought to be responsible for SC, while the Re-O layer (or F-doped) provides charge carriers, quite alike to that of cuprates. The layered crystal structure for this system is depicted in Fig. 1, with a schematic unit cell and Fe-As layer. The intra-layer and inter-layer chemical bonding is of covalent and ionic character, respectively. For the undoped ReFeAsO, a spin density wave (SDW) order seemed confirmed at low temperatures by neutron diffraction studies [11, 12]. Previous experiments indicated that by F-doping at the oxygen site, SC could be realized in these compounds by suppressing the SDW order and be further enhanced by the chemical pressure shrinking the crystalline lattice [8]. A theoretical calculation suggested that the enhancement of the density of states either by carrier doping or pressure is playing a key role for the appearance of SC [13, 14]. From this point of view, producing oxygen vacancies instead of F-doping which will create more electron carriers and favor the lattice shrinkage, would be a more efficient approach for the realization of SC in this class of materials. Based on these considerations, we have succeeded in preparing the ReFeAsO$_{1-\delta}$ superconductors with Re = Sm, Nd, Pr, Ce, and La by high pressure (HP) synthesis.

Here we report the SC and phase diagram in the ReFeAsO$_{1-\delta}$ family, induced by oxygen vacancies instead of F-doping, making this system much simpler for both experimental and theoretical studies. The tunable oxygen content that leads to the occurrence of SC in these compounds strongly resembles the situation in cuprates. The good metallic behavior, high-$T_c$ and high critical field of these materials is likely to make them another competitive candidate for high temperature SC applications.

A series of ReFeAsO$_{1-\delta}$ superconductors were prepared by HP synthesis as reported [9], and the nominal composition with $\delta = 0.15$ was found to be optimal for both higher $T_c$ and purer single-phase formation under our current synthesis conditions. The resistivity for all samples was measured by the standard four-probe method with indium-press electrodes. In Fig. 2, the temperature dependence of resistivity for the nominal ReFeAsO$_{0.85}$ (Re = Sm, Nd, Pr, Ce and La) samples is plotted together for comparison. All curves show clear superconducting zero resistivity transitions ($T_c$(zero)) at low temperatures, and the highest onset transition temperature ($T_c$(onset)) is 55.0 K for SmFeAsO$_{0.85}$ with a $T_c$(zero) at 52.8 K. For Re = Nd, Pr, Ce, and La, the $T_c$(onset) is 53.5 K, 51.3 K, 46.5 K and 31.2 K, respectively. All these superconductors show good metallic behavior from $T_c$ up to 300 K.

The corresponding DC magnetic susceptibility for these ReFeAsO$_{0.85}$ samples were measured under 1 Oe after both zero-field-cooling (ZFC) and field-cooling (FC) in a Quantum Design MPMS XL-1 system, and the results are plotted in Fig. 3. The sharp diamagnetic superconducting transition indicates the good sample quality, and the high superconducting shielding volume fraction reveals the bulk nature of these superconductors. For the SmFeAsO$_{0.85}$ sample, the 10% and 90% magnetic transitions are at 54.6 K and 52.6 K, respectively, as determined from the FC curve and the onset of this magnetic transition is at 55.0 K, *i.e.* the same as the onset of the resistivity transition.

The crystal structures for all ReFeAsO$_{1-\delta}$ samples prepared under high pressure or ambient pressure (AP) were characterized by powder X-ray diffraction (XRD) on an MXP18A-HF type diffractometer with Cu-$K_\alpha$ radiation from 20° to 80° with a step of 0.01°. All XRD patterns indicate the LaFeAsO structure being the main phase, while minor impurities (identified to be non-superconducting by-products) appeared when oxygen vacancies were introduced (for the same reason as explained in [9].) Some of the typical XRD patterns are plotted in Fig. 3 (a). The diffraction peaks of HP samples become broader and are shifted to higher angles in comparison with AP samples, indicating the smaller grain size with defects and shrunken lattice parameters for the HP samples. To reveal the relationship between the increasing $T_c$ and the substitution using smaller Re element, $T_c$ vs. lattice parameter '$a$' for ReFeAsO$_{0.85}$ is plotted in Fig. 3 (b) for different Re elements. In Fig. 3 (c), a series of NdFeAsO$_{1-\delta}$ samples with different oxygen concentration were studied and the $T_c$ vs. lattice parameter '$a$' is plotted in the phase diagram. Here the adoption of '$a$' instead of the nominal $\delta$ gives a better understanding for the phase diagram, since empirically there is a roughly linear relationship between the lattice parameter and the actual doping level [15], while currently the actual δ cannot be determined precisely from the polycrystalline samples. In Table 1, the lattice parameters and critical temperatures for the HP synthesized ReFeAsO$_{0.85}$ and AP synthesized ReFeAsO are given for comparison.

As seen in Fig. 3 (b), a shrinkage of about 3.1% in the Fe-As plane (for *a*-axis) was observed from Re = La to Sm for the nominal ReFeAsO$_{0.85}$, and with this increase of the inner chemical pressure on the Fe-As plane, the onset $T_c$ was dramatically enhanced up to 55 K from 31.2 K. For all five ReFeAsO$_{1-\delta}$ systems studied, in the under-doped region, SC emerges from the suppression of SDW state by a light vacancy doping. All these results indicate that the inner chemical pressure has a significant enhancing effect on the SC state as long as the competition between SC and SDW states is over [8].

Since the increase of oxygen vacancy will lead to both increase of the electron carrier density and the chemical pressure on the Fe-As plane, to investigate their respective influence on the SC state, a detailed study on the NdFeAsO$_{1-\delta}$ compound for different oxygen vacancy content was performed. For pure NdFeAsO, the SDW state at low temperatures was observed as previously reported [16]. By a slight doping of oxygen vacancies, the SDW was quickly suppressed and SC occurred immediately with an abrupt increase of $T_c$. The $T_c$(onset) increased to a maximum of 53.5 K when '*a*' was reduced to 3.943(7) Å from 3.965(3) Å of the undoped sample (about 0.55%). However, with further doping increase, the $T_c$ started to decrease slowly, instead. Comparing it with SmFeAsO$_{0.85}$ which has a smaller '*a*' and a higher $T_c$, the chemical pressure effect on the Fe-As plane is not likely to saturate. Therefore, the non-monotonic effect of the doping increase should be due to a competing factor: Over-doping of electron carriers should be the key factor that hinders the further increase of $T_c$.

To reiterate, we have succeeded in synthesizing a new family of iron-based arsenic-oxides superconductors using oxygen vacancies instead of fluorine-doping. We believe they will be the more promising prototype compounds for studying this new class of high-$T_c$ superconductors.

Note added at revision: After submitting of the present paper, a large physical pressure effect was reported in the F-doped La-based compound with the $T_c$ increased from 26 K to 43 K by a pressure of 3 GPa [17], which seems to be consistent with our results.

Acknowledgements:

We sincerely thank Prof. L. Yu, Z. Fang and X. Dai for lots of valuable discussions and Mrs. Shun-Lian Jia for her kind help in resistivity measurements. This work is supported by the Natural Science Foundation of China (NSFC, No. 50571111 & 10734120) and the 973 program of China (No. 2006CB601001 and No. 2007CB925002). We also acknowledge the support from EC under the project COMEPHS TTC.

Corresponding Authors:

Zhi-An Ren: renzhian@aphy.iphy.ac.cn

Zhong-Xian Zhao: zhxzhao@aphy.iphy.ac.cn

Figure and table captions:

Figure 1: The crystal structure for the ReFeAsO compound (space group: P4/nmm).

Figure 2: The temperature dependence of resistivity for the nominal ReFeAsO$_{0.85}$ samples synthesized by HP-method.

Figure 3: The temperature dependence of DC susceptibility (with FC and ZFC curves) for the nominal ReFeAsO$_{0.85}$ samples synthesized by HP method.

Figure 4: The comparison of typical XRD patterns for some ReFeAsO$_{1-\delta}$ samples synthesized by HP and AP method (upper solid line: HP samples for the nominal ReFeAsO$_{0.85}$; lower dashed line: AP samples for the undoped ReFeAsO; vertical bars: calculated diffraction peaks based on the undoped PrFeAsO), impurity peaks are marked with '*' (a); the lattice parameter dependence of $T_c$ for the nominal ReFeAsO$_{0.85}$ samples synthesized by HP method, the solid line is for guiding (b); the lattice parameter dependence of $T_c$ for the nominal NdFeAsO$_{1-\delta}$ samples synthesized by HP method (AP for the $\delta = 0$ sample), the solid line is for guiding (c). Note: The SDW state was confirmed in La-based compound [11, 12], while it should be checked for the present system.

Table 1: Some experimental data for the nominal ReFeAsO$_{0.85}$ superconductors and the undoped ReFeAsO compounds.

Table 1.

| Re | a (Å) | c (Å) | $T_c$(onset-R) (K) | $T_c$(zero-R) (K) | $T_c$(onset-M) (K) |
|---|---|---|---|---|---|
| Sm (HP) | 3.897(6) | 8.407(1) | 55.0 | 52.8 | 55.0 |
| Sm (AP) | 3.933(5) | 8.495(3) | | | |
| Nd (HP) | 3.943(7) | 8.521(8) | 53.5 | 50.9 | 51.0 |
| Nd (AP) | 3.965(3) | 8.572(3) | | | |
| Pr (HP) | 3.968(2) | 8.566(1) | 51.3 | 43.2 | 48.2 |
| Pr (AP) | 3.985(8) | 8.600(3) | | | |
| Ce (HP) | 3.979(7) | 8.605(5) | 46.5 | 41.0 | 42.6 |
| Ce (AP) | 3.998(1) | 8.652(6) | | | |
| La (HP) | 4.022(2) | 8.707(1) | 31.2 | 25.9 | 28.3 |
| La (AP) | 4.033(5) | 8.739(1) | | | |

Note: The HP samples are for the nominal composition of ReFeAsO$_{0.85}$, and the AP samples are for the nominal composition of ReFeAsO. And here we also note that $T_c$(onset) ~ 36 K in LaFeAsO$_{1-\delta}$ has been achieved by HP synthesis but the sample quality is still not good enough.

Figure 1.

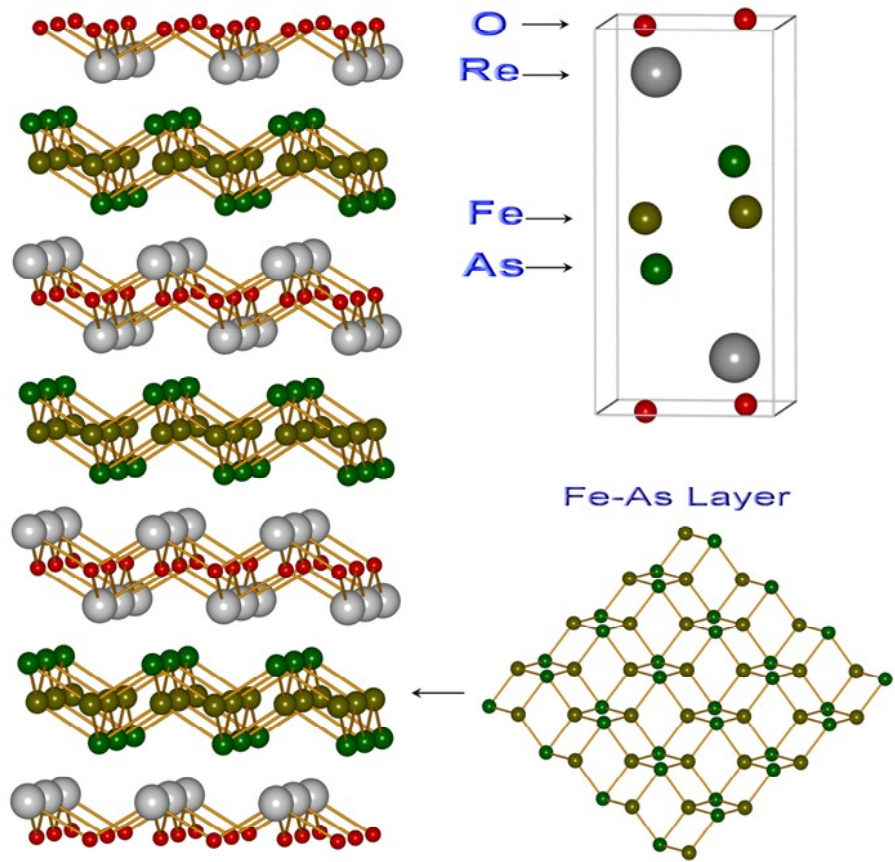

Figure 2.

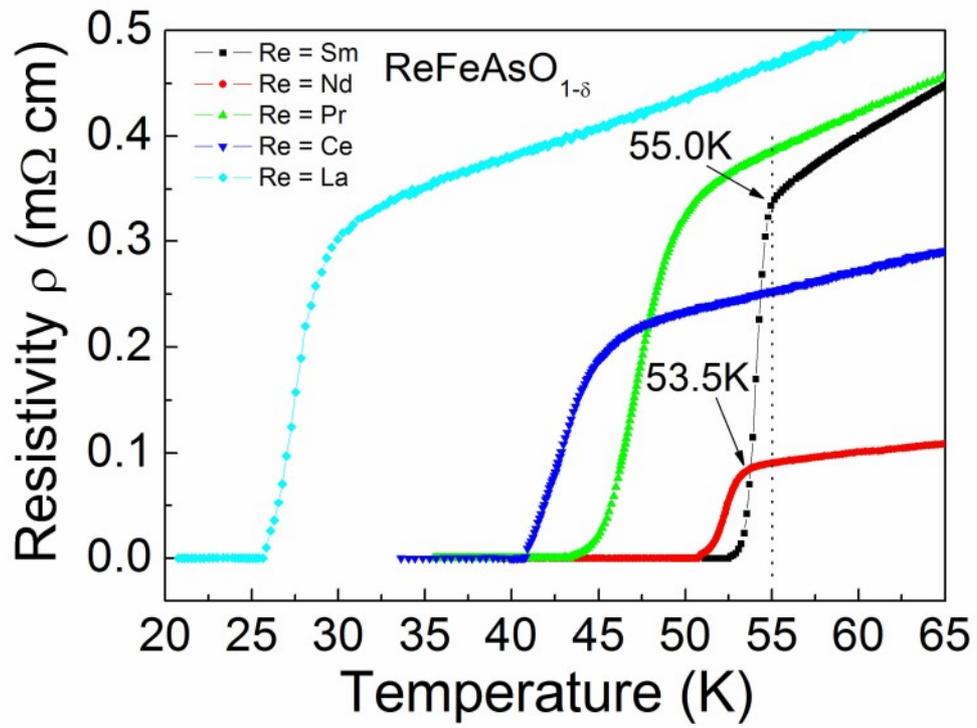

Figure 3.

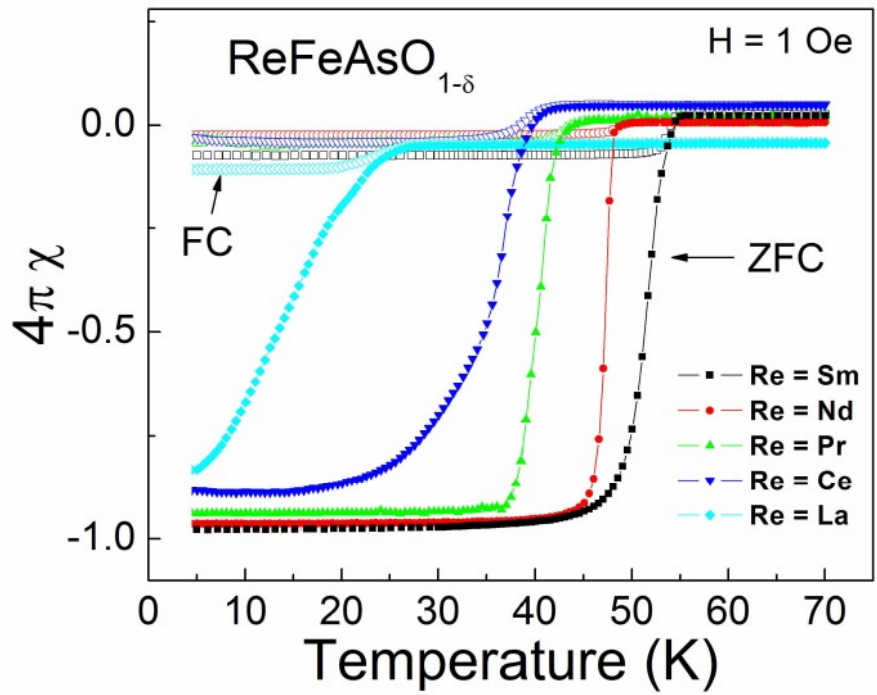

Figure 4.

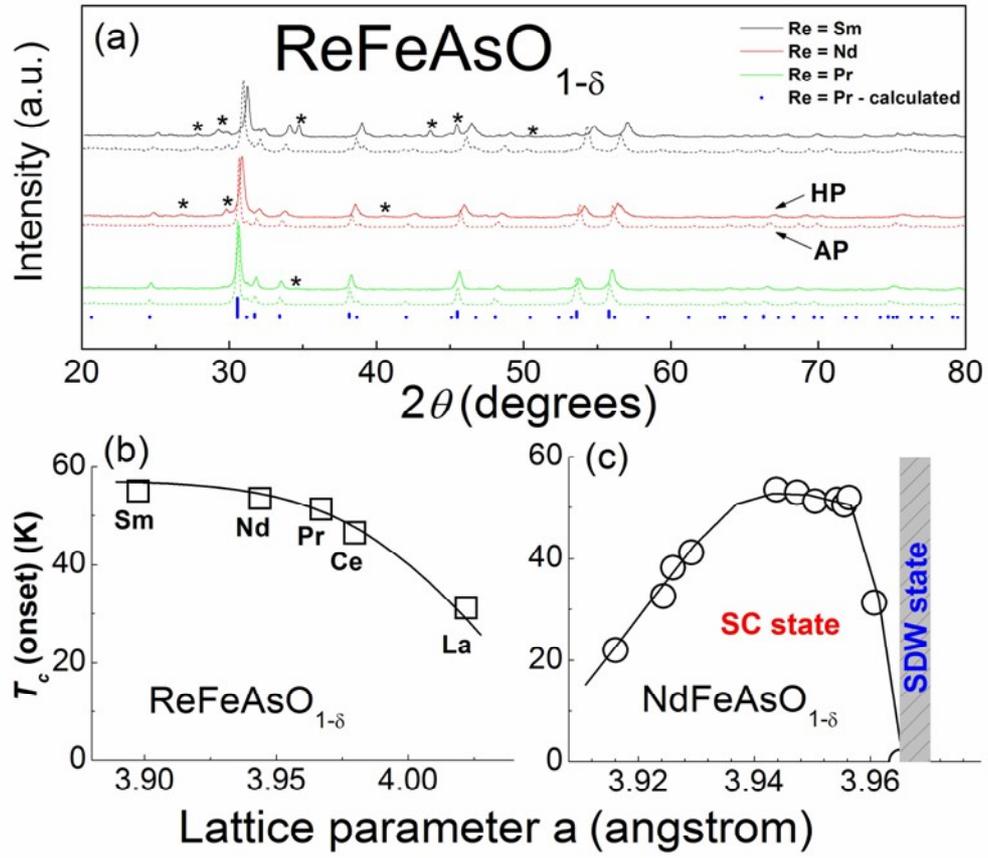